\documentclass[conference]{IEEEtran}
\IEEEoverridecommandlockouts
\usepackage{cite}
\usepackage{amsmath,amssymb,amsfonts}
\usepackage{algorithmic}
\usepackage{graphicx}
\usepackage{textcomp}
\usepackage{xcolor}
\usepackage{url}
\def\BibTeX{{\rm B\kern-.05em{\sc i\kern-.025em b}\kern-.08em
    T\kern-.1667em\lower.7ex\hbox{E}\kern-.125emX}}
\begin{document}

\title{always\_comm: An FPGA-based Hardware Accelerator for Audio/Video Compression and Transmission}

\author{\IEEEauthorblockN{Rishab Parthasarathy\textsuperscript{\textsection}}
\IEEEauthorblockA{\textit{MIT} \\
rpartha@mit.edu}
\and
\IEEEauthorblockN{Akshay Attaluri\textsuperscript{\textsection}}
\IEEEauthorblockA{\textit{MIT} \\
akshay\_a@mit.edu}
\and
\IEEEauthorblockN{Gilford Ting\textsuperscript{\textsection}}
\IEEEauthorblockA{\textit{MIT} \\
gting@mit.edu}}

\maketitle

\begin{abstract}
We present a design for an extensible video conferencing stack implemented entirely in hardware on a Nexys4 DDR FPGA, which uses the M-JPEG codec to compress video and a UDP networking stack to communicate between the FPGA and the receiving computer. This networking stack accepts real-time updates from both the video codec and the audio controller, which means that video will be able to be streamed at 30 FPS from the FPGA to a computer. On the computer side, a Python script reads the Ethernet packets and decodes the packets into the video and the audio for real time playback. We evaluate this architecture using both functional, simulation-driven verification in Cocotb and by synthesizing SystemVerilog RTL code using Vivado for deployment on our Nexys4 DDR FPGA, where we evaluate both end-to-end latency and throughput of video transmission.
\end{abstract}


\section{Introduction and Motivation}
\begingroup\renewcommand\thefootnote{\textsection}
\footnotetext{ Equal contribution}
\endgroup
As video conferencing has become increasingly common post-COVID, video streaming has become an extremely costly workload for traditional computers. In response, we present always\_comm, a design for an FPGA-based video conferencing and streaming system capable of streaming video in real time, either to another FPGA or an online endpoint. At the highest level, our specialized device should be able to accept video data directly from a camera and speaker and perform real-time signal processing and compression so that it can be streamed over a network connection.

Specifically, we demonstrate that an FPGA can calculate video codecs and audio in parallel with networking protocols, enabling efficient computation of the video conferencing stack. With core modules calculating the Motion JPEG codec for video and communicating audio, image data can be streamed directly from BRAM to a nearby laptop connected over Ethernet using our UDP networking module. This means that our project can be used for video conferencing (if it’s a one person to one person connection) as it can stream video at 30 FPS. Overall, we demonstrate that with limited memory, an FPGA-based implementation can achieve extremely high throughput in video conferencing, achieving the same as with much bulkier, more hardware intensive implementations on classical compute like our computers.

There are a number of significant challenges that makes this project meaningful to do. First, while an FPGA may be better able to parallelize the computation of the DCT, we must also be extremely careful to not overuse DSPs. On top of that, run-length encoding requires variable length bitstreams, which is difficult to manage on a traditional FPGA module which expects fixed size outputs. Overall, the Ethernet PHY on the Nexys4 DDR boards is also not very well documented, so there are significant challenges with transmitting over this connection with properly formed packets.

As such, we set the goals of the project like so:
\begin{enumerate}
    \item [1.] \textbf{The Commitment (MVP):} Able to send arbitrary packets over Ethernet and able to demonstrate a minimal compression scheme that can be sent over the Ethernet and decompressed to a larger size.
    \item [2.] \textbf{The Goal:} Able to send arbitrary packets over UDP and able to demonstrate that there is a functioning M-JPEG pipeline.
    \item [3.] \textbf{Stretch Goal:} Fully functioning UDP and M-JPEG protocol that operate and can be processed at 30 FPS.
 \end{enumerate}

 As such, the current state of our system contains a UDP module that sends over the Ethernet at 87 FPS. On top of that, through Wireshark traces and Python scripts, we have found that we can use Huffman encoding to compress and decompress packets with at least a compression ratio of 8-12. We have also implemented a successful testbenched version of the DCT, which completes the M-JPEG protocol. All portions of our codebase have been verified using functional verification with Cocotb, along with on-silicon testing using a Nexys4 DDR FPGA.

\section{System Description}
At a high level, the system consists of three key modules: the audio module, the video codec, and the networking interface. 

First, the audio module consists of a SPI controller reading from a microphone at 8 kHz, which is written to our other modules through the network controller. We provide a simple block diagram below in Fig. \ref{fig:audioblock}.

\begin{figure}
    \centering
    \includegraphics[width=\linewidth]{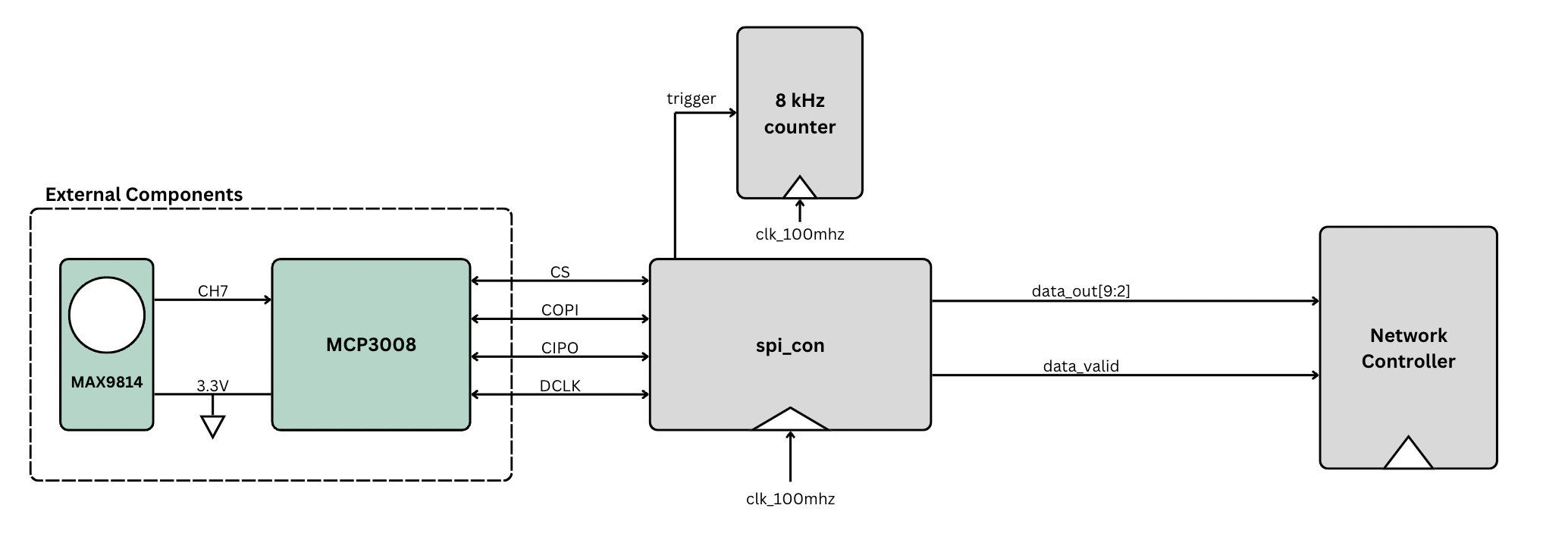}
    \caption{Block diagram for audio module}
    \label{fig:audioblock}
\end{figure}

\begin{figure*}
    \centering
    \includegraphics[width=\linewidth]{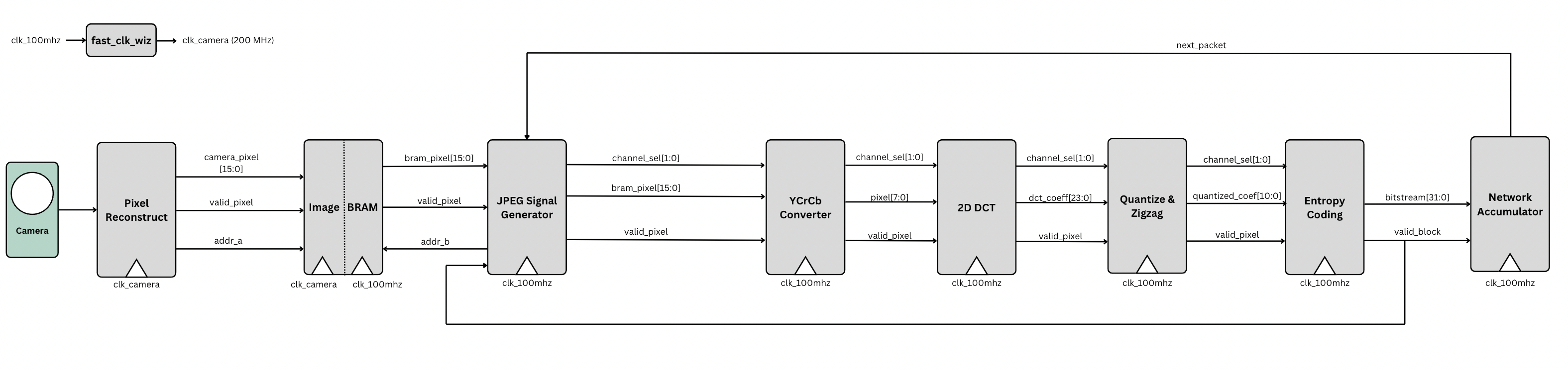}
    \caption{Block diagram for video module}
    \label{fig:videoblock}
\end{figure*}

Next, the video codec module is a compressor for video streaming at 30 FPS, which otherwise would have an extremely large memory footprint and thus be infeasible to transmit efficiently. Specifically, the video codec is connected to a frame buffer, which retains the frame due to the camera being in a different clock domain than the rest of the code. First, the image is input into a JPEG signal generator that retrieves the appropriate section of the image, which is converted into YCrCb form, an alternative to RGB that encodes luminance and chrominance. This YCrCb form is input into the video codec module, which in this case implements the MJPEG codec. This MJPEG codec has each frame encoded as a separate JPEG image, which is encoded using the Discrete Cosine Transform (DCT), a quantization, and Entropy Coding. The entropy coding consists of separate run length encoding and Huffman encoding. Finally, this video codec output is streamed into a network accumulator. The video codec protocol is described in more depth in Section 3 of this report, and a high level block diagram can be seen in Fig. \ref{fig:videoblock}.

\begin{figure}
    \centering
    \includegraphics[width=\linewidth]{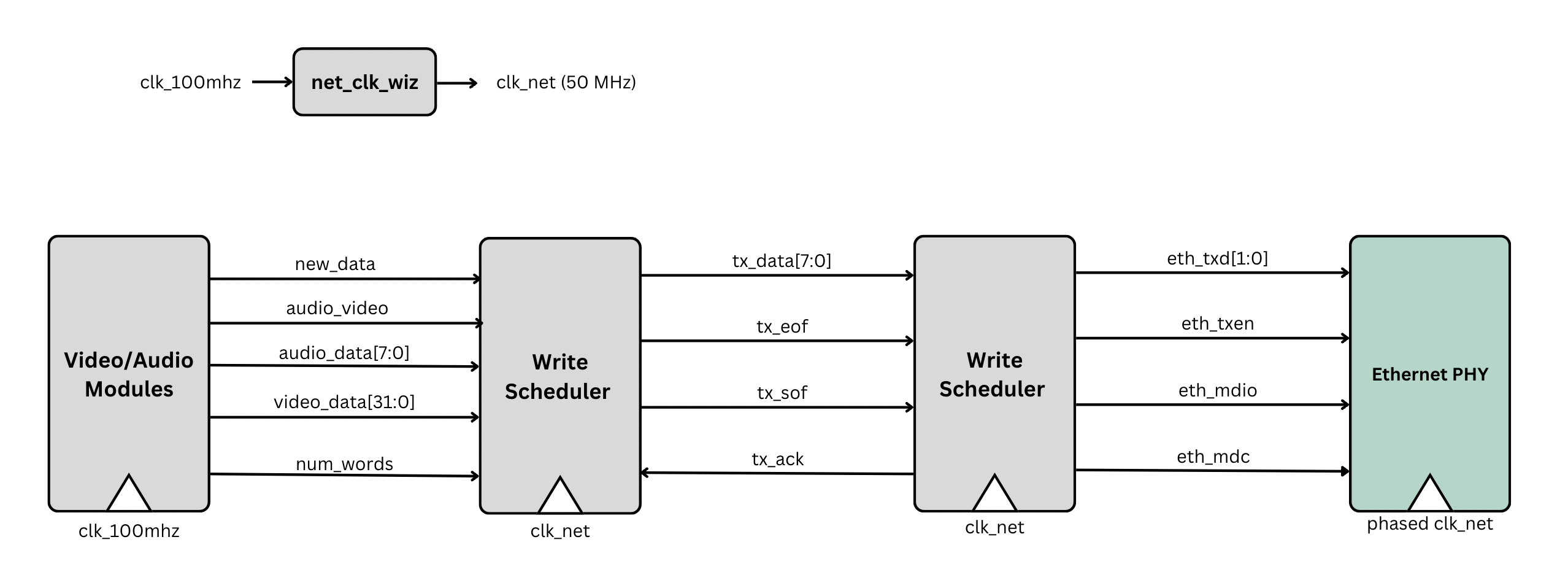}
    \caption{Block diagram for networking module}
    \label{fig:network_block}
\end{figure}

\begin{figure}
    \centering
    \includegraphics[width=0.8\linewidth]{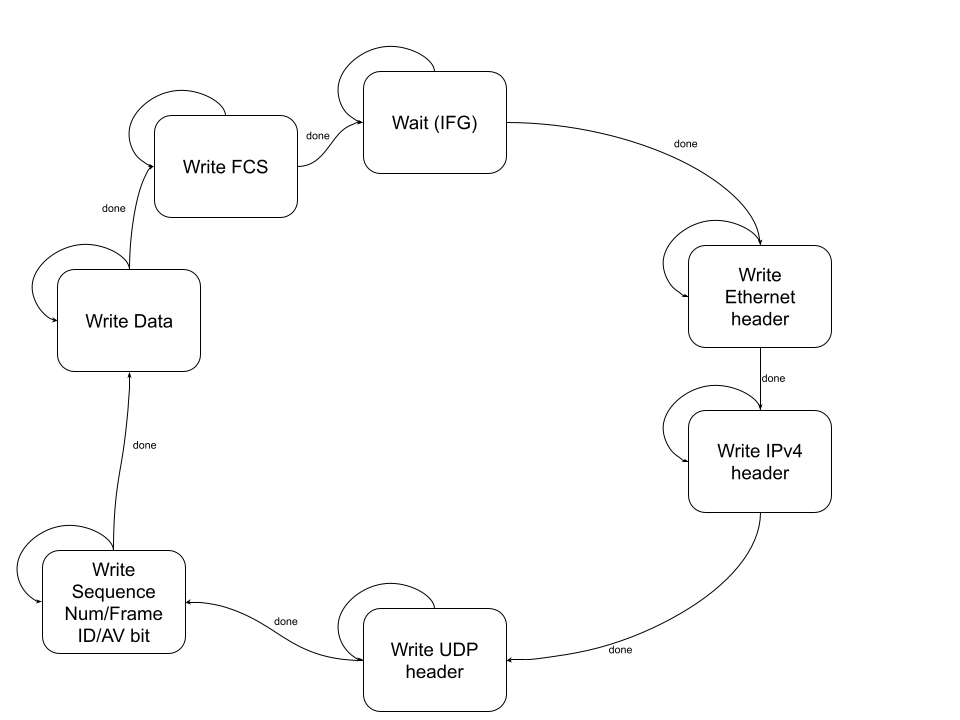}
    \caption{FSM for networking module writing to Ethernet PHY}
    \label{fig:ethfsm}
\end{figure}

Finally, the networking module is a finite state machine that reads in data from the video and audio registers, performs necessary computations to generate headers and checksum, and outputs this information to the Ethernet PHY. The Ethernet PHY on our Nexys4 DDR board follows the Reduced Media Independent Interface (RMII) protocol, meaning that we have to send it information 2 bits per cycle of a 50 MHz clock (100 Mbps). To this end, we have a transmit scheduler that transitions our networking module between its various states, each of which involves writing a different component of our packet to the PHY. The block diagram is provided in Fig. \ref{fig:network_block}. The FSM is also defined in Fig. \ref{fig:ethfsm}. 

\section{Specific Design}
\subsection{Audio Module}
Because our video codec module is significantly more complex than most audio codec schemes (and deals with much more data), we use a simple audio transmission module, reading out the audio signal from a MAX9814 microphone and storing it in an audio BRAM block. This data is read once the Ethernet module requires audio data, when in the process of assembling a packet. This reads 8 bits of data at 8 ksps, which is fed into a register that is read 10 times a second, meaning that the register is 8 Kb. We use a register and not a BRAM because this makes synchronizing signals for the networking module significantly easier. This communicates with the speaker over SPI.
\subsection{Video Codec}
The video module connects with an OV5640 camera clocked at 12.5 MHz using a traditional frame buffer, which is then piped into the MJPEG specification, which processes each frame of the video separately using the JPEG codec. We describe the detailed implementation of the specification below.
\subsubsection{YCrCb Subsampling and JPEG Signal Generator}
The first step of the MJPEG specification is subsampling the channels in YCrCb. Because the luminance channel is more important than the chrominance channels for human eyesight, the standard subsampling scheme is 4:2:0, which means that all luminance values are retained, but each 2x2 grid of chrominance values is averaged. This module also rescales the range of the values from [0, 255] to [-128, 127], as the transformation used next is more stable when values are centered around 0 rather than solely positive. 

To implement this subsampling, we divide the image into 16x16 superblocks, as each DCT operates on 8x8 chunks. Each superblock thus contains 4 blocks in the non-subsampled Y channel and 1 block in each of Cr and Cb, amounting to 6 blocks in total. We note that because each superblock thus has 6 chunks of size 64, which have DCT coefficients of at most 11 bits, we find that two superblocks fit in each Ethernet packet. We thus trigger our JPEG signal generator to traverse two superblocks each time a new video packet is requested by the network accumulator. We also traverse the next block for DCT only after the previous DCT is fully done being JPEG compressed, as this simplifies synchronization and still easily meets timing as discussed in the Evaluation section.
\subsubsection{2D DCT}
The second step of the MJPEG specification is computing the 2D discrete cosine transform (DCT) of the subsampled Y, Cr, and Cb channels, which isolates the different frequency components for later compression.

The 2D DCT follows the formula
\begin{equation}
    G_{x,y} = \frac{1}{4} \sum_{i = 0}^7 \sum_{j=0}^7 g_{i, j} \cos\left(\frac{(2i + 1)x \pi}{16}\right) \cos\left(\frac{(2j + 1)y \pi}{16}\right)
\end{equation}

\begin{figure}
    \centering
    \includegraphics[width=\linewidth]{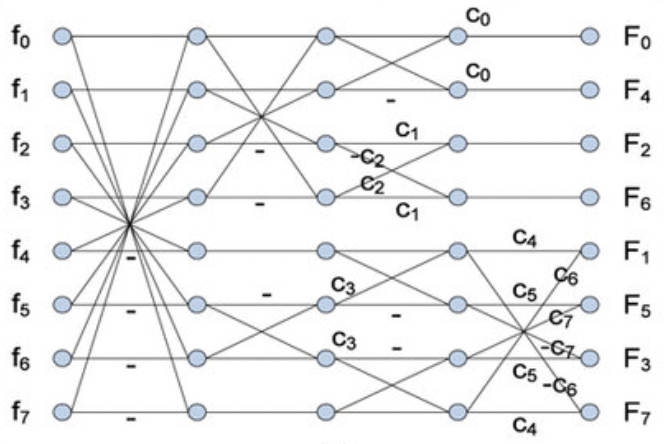}
    \caption{The butterfly algorithm for the DCT \cite{b1}.}
    \label{fig:butterfly}
\end{figure}

Analyzing this equation, we note that the 2D DCT is fully separable into a 1D DCT on the rows and a 1D DCT on the columns, which we leverage in our implementation. We begin by implementing a 1D DCT module, which calculates the 1D DCT on a series of 8 values. This module accepts a series of 8 values every cycle and outputs the relevant outputs 5 cycles later. To pipeline this module, a butterfly-based algorithm was utilized as displayed in Fig. \ref{fig:butterfly}.
The butterfly algorithm splits the DCT into antisymmetric and symmetric terms, which are then multiplied by the cosine coefficients, reducing the number of multiplications that must be performed by 30\% \cite{b1}. This algorithm was implemented using fixed point arithmetic, with the output of the DCT module having 5 extra bits after the decimal point as compared to the input. Fixed point arithmetic was chosen to reduce the DSP load from the discrete cosine transform, as this is the most DSP-intensive part of the project.

Then, the 2D DCT is implemented as an FSM that calls the 1D DCT, which first loads in 64 serialized values over 64 cycles. After that, the 1D DCT is called on each row of the loaded image, and then, the 1D DCT is called on each column of the outputs of the 1D DCT. This output is then serialized and outputted over 64 cycles. We choose to serialize input and output, using an FSM, because this reduces the load on registers without introducing a large amount of dead time, as discussed in Evaluation.

\subsubsection{Quantization}
The quantization step accepts the outputs of the DCT module and scales the output by dividing by some integer and rounding the result. To do this, we use a standard quantization matrix proposed in the original JPEG specification as the 50\% quality threshold \cite{b2}. The quantization module accepts serialized inputs and outputs serialized outputs, keeping an internal counter from 0 to 63 that keeps track of which quantization value to use. This module is also implemented using fixed-point arithmetic by keeping track of the inverses of the divisors and using multiplications rather than divisions.

\subsubsection{Zigzag Readout}
\begin{figure}
    \centering
    \includegraphics[width=0.7\linewidth]{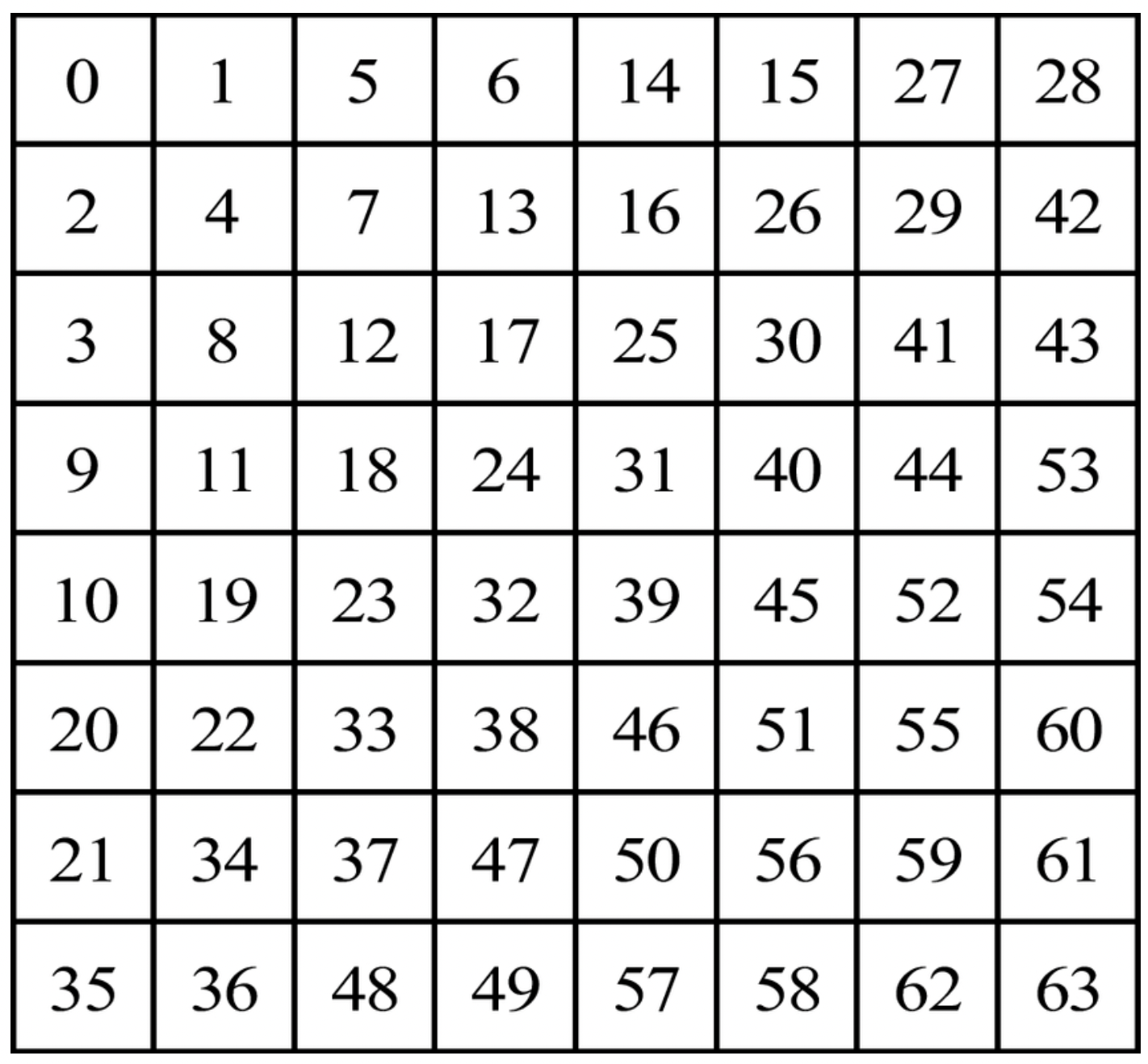}
    \caption{Zigzag ordering for reading out the quantized values \cite{b3}.}
    \label{fig:zigzag}
\end{figure}

After quantization, because of the DCT's tendency to encode large values in the upper left quadrant of the 8x8 grid, the values are read in a zigzag order to shift nonzero values to the beginning of the serialized sequence \cite{b3}. A diagram of the ordering is shown in Fig. \ref{fig:zigzag}.

To implement this ordering, the module accepts serialized inputs for 64 cycles before switching to outputting the serialized inputs in the correct order for 64 cycles. The output order is stored using a lookup table, which can be used to index into the inputs.
\subsubsection{Entropy Coding}
The entropy coding module receives as input a serial stream of 11-bit quantized DCT coefficients, in the order defined by the zigzag readout module. In accordance with the JPEG specification, we use three different methods to achieve a minimal-length output: a category-based value representation, run-length encoding, and Huffman encoding. Importantly, this module has \textit{variable-length} outputs, which will eventually be turned into an output bitstream by downstream modules.

Firstly, coefficient values are not transmitted exactly as-is; rather, a categorization scheme is used to minimize the number of bits needed to represent these values. Specifically, a given value $V$ has the category $C = \lfloor\log_2|V|\rfloor + 1$, and is transmitted as a $C$-bit value -- conversely, the values for a given category $C$ fall in the ranges $[1-2^C, -2^{C-1}]$ and $[2^{C-1}, 2^C-1]$. (The value 0 has the category 0 and is not represented by any bits.) The actual $C$-bit value that is sent depends on the \textit{sign} of $V$ -- if $V > 0$, $V$ is sent, and if not, $\sim V$ (the logical negation of $V$) is sent instead. For example, the value category $C = 3$ contains the values $-7, -6, -5, -4, 4, 5, 6,$ and $7$, and the bit representation of these values are $000, 001, 010, 011, 100, 101, 110$, and $111$. Under this representation, the number of bits used to represent a value in the final bitstream is related to its magnitude, which is beneficial since many of our DCT coefficients are often small numbers -- specifically, this avoids sending large numbers of bits for small negative numbers (a regular 11-bit two's complement representation of $-1$, for example, uses 11 bits).

Run-length encoding takes advantage of the fact that many of the values in the resultant DCT coefficient block are equal to 0, and significant compression can be achieved by simply counting the number of zeroes preceding any nonzero coefficient. To achieve this, an FSM structure is used, where the module only transmits information downstream if it receives a nonzero coefficient and simply increments a \texttt{zero\_run\_length} counter otherwise. There are also two signals transmitted for edge cases -- \texttt{zero\_run\_length} is of width 4, so it has a maximum value of 15. Thus, when the 16th zero in a row is processed, a special \texttt{<ZRL>} code representing 16 consecutive zeroes is transmitted. Additionally, the last coefficient in the block is followed with the end-of-block code \texttt{<EOB>} (if there are preceding zeroes and the last coefficient is zero, \texttt{<EOB>} is sent directly).




Thus, each fundamental chunk of information is a tuple of 3 numbers: $R$, the length of the run of zeroes preceding the nonzero coefficient, $N$, the minimum number of bits required for the value, and $V$, the value of the coefficient. Crucially, $R$ and $N$ are \textit{fixed-length} quantities, while $V$ is a \textit{variable-length} quantity. As long as the values are sent to a downstream reader in this order, the coefficients can be easily recovered. For our specific protocol, one reads 4 bits from the stream to get $R$, another $4$ bits to get $N$, and the next $N$ bits to get $V$. 

Lastly, the application of Huffman encoding comes from the observation that the set of possible pairs of $R$ and $N$ is relatively small, and often repeat -- for example, the sequence (0, 0, 0, 7) and the sequence (0, 0, 0, 4) have the same values of $R = 3$ and $N = 3$. Thus, a Huffman encoding scheme can be used to compress the ($R$, $N$) pairs losslessly. We use a lookup table stored in read-only-memory (ROM) to translate a code \textit{index}, the concatenation of $R$ and $N$, into a \textit{codeword}, a prefix-free binary code with variable length (between 2 and 16 bits). As per the JPEG standard, each combination of channel type (Y vs Cr/Cb) and coefficient type (DC vs AC, top left coefficient vs all other coefficients) get its own set of Huffman codes -- we use the values listed in the tables of the JPEG standard \cite{b7}.

At the end of this module, we have a \texttt{codeword} Huffman code output, a \texttt{coeff} value represented using the categorical scheme, and other outputs to keep track of how many bits in these values carry meaning. Huffman codes are left-aligned, and coefficient value are right-aligned. To combine these outputs and serialize them properly, we pass them into the next module.
\subsubsection{Data Alignment}
\begin{figure}
    \centering
    \includegraphics[width=0.8\linewidth]{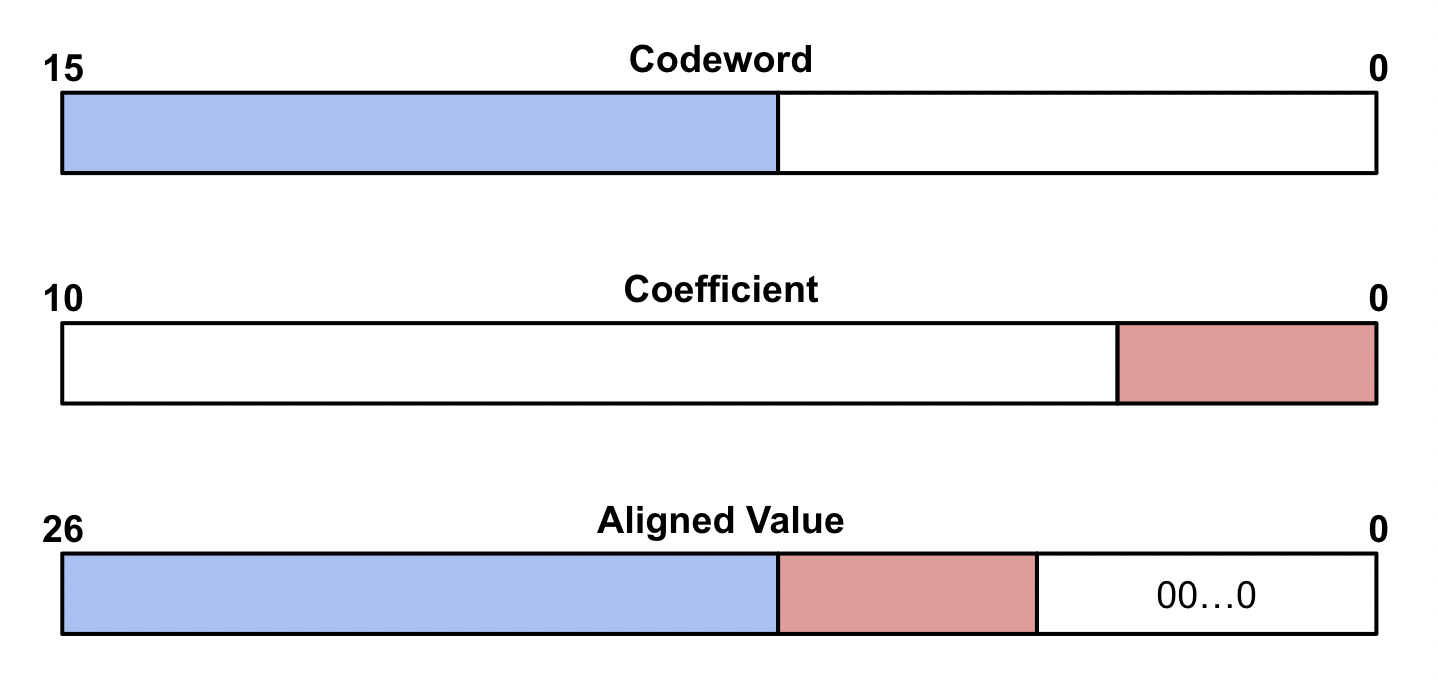}
    \caption{Alignment of compressed coefficient information}
    \label{fig:alignment}
\end{figure}

The data alignment module combines \texttt{codeword} and \texttt{coeff} into a 27-bit value \texttt{aligned\_value} such that the relevant bits are left-aligned and right next to each other, with the rest of the variable being zeroes. (See Figure \ref{fig:alignment}.) We also keep track of how many bits in \texttt{aligned\_value}, carry meaning and both values are passed into the next module for serialization.
\subsubsection{Serialization}
The ultimate goal of the serialization module is to remove all ``dead" bits that result from fitting variable-length codes into fixed-length variables. This is achieved by reading the \texttt{aligned\_value}s into a buffer every time valid data is available, and outputting a 32-bit word every time the buffer accumulates more than 32 bits. In this manner, the variable-length compressed coefficients are turned into a stream of bits divided into 32-length chunks. These chunks are put into a video register for the networking module. There is no danger of the buffer internal to the module overflowing infinitely, as the maximum number of bits arriving per cycle is less than the number of bits in the output (27 $<$ 32).
An important implementation detail is that when the overall length of the stream may not be divisible by 32, the last 32-bit chunk will not be fully filled. We pass along an \texttt{end\_of\_block} signal with the previous variables (pipelined appropriately, of course) such that this module will know when to output an incomplete chunk.

\subsection{Integration}
Because the network is able to reorder and drop packets, we make a few key changes to the integration between the modules. Firstly, within each packet, we send two ``superblocks", with each superblock comprised of 4 Y blocks, one subsampled Cr block, and one subsampled Cb block. The bits are aligned to the boundary between bitstream packets, and are thus outputted in aligned chunks of size 32 bits, such that each packet can be independently decoded into two 16x16 blocks of the image. In each packet, we also send a position number -- this is simply the index of the two-superblock chunk when iterating over the blocks of the image in row-major order. Because 180 (the height of our output video) is not divisible by 16, we pad to the nearest multiple of 16 using zero-valued pixels. Furthermore, we encode the DC coefficients as their raw values instead of taking the difference from the previous block -- this removes dependencies between blocks and allows us to deal with reordering packets.

\subsection{Networking Interface}
As defined above, the networking interface is managed by a scheduler that maintains an FSM, where each state involves writing a different portion of our Ethernet packet to the Ethernet PHY. The scheduler remains in each state for the appropriate amount of cycles to fully transmit the relevant data, at the 2 bit/cycle rate defined by the PHY. These values can be found in Table \ref{table:packet}. Note that the networking module runs on a 50 MHz clock to accomodate for the Ethernet PHY's specifications \cite{b7}, generated from a clock wizard. Now, we discuss the individual components of the state machine, as described in the table, and in the following subsections.

\begin{table}[h!]
\centering
\begin{tabular}{|l|c|c|}
\hline
\textbf{Field}         & \textbf{Size}                & \textbf{Cycles} \\ \hline
Preamble               & 8 bytes                      & 32 cycles       \\ \hline
Ethernet Header        & 14 bytes                     & 56 cycles       \\ \hline
IPv4 Header            & 20 bytes                     & 40 cycles       \\ \hline
UDP Header             & 8 bytes                      & 32 cycles       \\ \hline
Audio/Video            & 1 byte                       & 4 cycles      \\ \hline
Sequence Number        & 1 byte                       & 4 cycles      \\ \hline
Data                   & Up to 1470 bytes             & Up to 5880 cycles     \\ \hline
FCS                    & 4 bytes                      & 16 cycles       \\ \hline
Interframe Gap         & -                            & 48 cycles       \\ \hline
\end{tabular}
\vspace{1em}
\caption{Packet Fields and Corresponding Sizes and Cycles}
\label{table:packet}
\end{table}

\subsubsection{Preamble}
In accordance to the RFC 1972 spec, we prepend our Ethernet packets with an 8 byte preamble, consisting of 7 bytes of \texttt{0x55}, and one byte of \texttt{0xD5}.

\subsubsection{Ethernet/IPv4 Headers}
We structure our Ethernet and IPv4 headers in accordance to the RFC 1972 and 791 specs. The specs define the header to consist of source and destination addresses, which we predefine to be local addresses. This is permissible due to our network consisting of a simple ether with only 2 clients. The rest of the headers are various pieces of metadata, such as TTL (time to live), etc.

\subsubsection{UDP Header}
We also have a predefined UDP header in accordance to RFC 768. This also has source and destination addresses, as well as other metadata. Under IPv4 the UDP checksum is optional, so we forgo it since we have the Ethernet checksum instead.

\subsubsection{Metadata}
Each data packet is accompanied with relevant metadata meant to expedite the reconstruction process. In particular, a 1-byte signal indicating audio vs. video data, and a 1-byte sequence number are prepended to the payload. The sequence number allows for reconstruction of out-of-order packets by introducing a canonical ordering before any reordering happens.

\subsubsection{Data}
We route the data from either the audio or video register, based on the current states of the registers. The audio register consists of 1024 8-bit registers, while the video register is 300 32-bit registers. The audio sampling module writes into these 8-bit registers directly, and the entropy coding module does the same for the video register. We select the video register size because the entropy coding pads its output words to 32 bits.

\subsubsection{PHY Handling}
The Ethernet PHY that is onboard the Nexys 4 DDR boards is the LAN8720A Small Footprint RMII 10/100 Ethernet
Transceiver. We provide it with input signals in accordance to the LAN8720A datasheet. In particular, we maintain inputs for an MDIO interface, including a 1.5625 MHz clock and MDIO signal. We also provide a reference clock that runs in -45º phase to the RTL networking logic. Finally, we maintain a transmit enable and reset signal for the PHY. The reset signal is held to an active low for the first 20000 cycles of the 50 MHz clock. The transmit enable signal is held high while sending data and headers, and is held low for the rest of the FSM's transitions.

We use counters of the register's occupancy as well as signals from the source modules to know when to send an Ethernet packet. In particular, we send an audio packet when the audio register reaches 800 8-bit registers full. We simply flush these bits into the packet and send it to the PHY. For video, we have a signal that is routed from the entropy coding module that indicates when a total of 12 blocks (2 superblocks) have been encoded. At this point, we flush the video bitstream from the registers into an Ethernet packet. Note that this encoding data has variable length — luckily, Ethernet packets can be anywhere from 64 to 1518 bytes in length, so we can easily fit the full range of values into our packets.

Upon sending of the whole data packet, we send a signal back to the JPEG signal generator module to start the codec stack.

\subsubsection{FCS}
We implement CRC32-IEEE 802.3 in order to generate a checksum for our packet. This implementation takes in a byte of data as it is received by the transmit scheduler, allowing for parallelization of the checksum calculation with writing of the packet.

\subsubsection{Overall State Machine}

This state machine is governed by two modules, \texttt{write\_scheduler} and \texttt{tx\_scheduler}. The first handles reading from data registers and handling the Headers, Metadata, and Data states. From these states, the \texttt{write\_scheduler} receives one bit per 100 MHz clock cycle, which is then converted to one byte per 4 Ethernet clock cycles at 50 MHz. Then, the \texttt{tx\_scheduler} handles the Preamble, FCS, and IFG states, and is responsible for direct interfacing with the PHY's inputs and outputs. During the relevant phases, it receives data at a rate of 1 byte per 4 cycles from the \texttt{write\_scheduler}, as discussed earlier, sending the data from the data registers to the Ethernet PHY.

\subsection{Reconstruction}
On the recipient device, we use a Python script with Wireshark to sniff the relevant Ethernet port and collect all packets. We write these packets to a \texttt{.csv} file, which is read by another script that performs the inverse of the video codec procedure. Audio is also similarly reconstructed. Due to poor concurrency of Python and execution speed of Python, we do not currently support a live video feed.

\section{Evaluation}

\subsection{Functional Verification}

All modules were functionally verified using the Cocotb simulation software and iverilog compiler. Functional verification was performed using a coverage-driven approach. First, target implementations were implemented using NumPy for each part of the MJPEG codec and the Ethernet transmission protocol. Then, edge cases and boundary cases were tested first, to ensure coverage of all boundary cases. Finally, random data points were autogenerated and verified functionally. In all cases, waveforms were also manually visually inspected to ensure that all computations were also qualitatively correct.

For the codec, we found that our use of fixed point arithmetic resulted in errors of less than 5\% relative to the higher precision NumPy implementation, within the margin of error. We also found that all aspects of run-length/entropy coding and Ethernet transmission were 100\% accurate compared to the functional specification.
\subsection{Latency and Throughput}
We qualify the latency and throughput of our overall system by measuring the time it takes to send a single packet of two superblocks. Each DCT takes around 150 cycles to complete (64 to process input, 20 for DCT and 64 to serialize output), and the quantize and zigzag modules take another 70 cycles. The run length encoding takes a final amount of at most 20 to 30 cycles to fully serialize the output. For each of the twelve blocks, this takes 240 cycles, while the final signal from the networking module takes 384 cycles to send a header and 1080 cycles to send the data values. The value of 1080 is taken empirically from the average package size over Wireshark of 135 bytes, meaning that overall, we send each packet in around 4400 cycles. Scaling that to 120 packets per frame and 100 million cycles per second, this means that we send frames at 180 FPS, which matches the observations found in packet decompression, where we observe a frame rate of 183 FPS.

We note that thus our decision to make our modules FSMs and drive DCTs sequentially for each packet does not impact our throughput significantly, as we are still able to process video far faster than needed for our goal of 30 FPS.

\subsection{DSP and BRAM usage}
In terms of DSP usage, the main use of DSPs in our project was the DCT module, as it computes a number of pipelined additions and multiplications. We find that the DCT module uses 44 of the 53 DSP blocks allocated to the design, with the majority of the rest being allocated to the YCrCb conversion. Given that the board has 240 DSPs, we find this to be a perfectly reasonable amount, as our butterfly based approach to the DCT saves a significant amount of computation.

On the other hand, for BRAM, we only use BRAM for the frame buffer, which is of size 1 Mb, and lookup tables for Huffman encoding, setting the camera registers, and the networking module. With this in mind, we are firmly under the threshold for BRAMs. We note that a reasonable extension of our project is to send the audio data and video data into FIFOs built on top of BRAMs as opposed to the current register setup, as this is the more natural solution on FPGAs. However, for the sake of reducing complexity and synchronization of signals, we choose to use registers as we have more than enough logic slices to account for this usage. In fact, we only use 9000 LUTs and 20000 FFs when each of 15000 logic slices either provides 6 LUTs or 8 FFs, meaning we still have a large amount of distributed RAM left.

\subsection{Synthesization and Timing Constraints}
In its current state, our system synths and deploys successfully with a final WNS of 0.255 on the Nexys4 DDR FPGA. We are able to send Ethernet packets containing valid video and audio data to a connected laptop, on which we can run a reconstruction script to generated decoded video and audio data.

Our entire stack's timing is based on the specification set by the Ethernet PHY, which is a data transmission of 2 bits every cycle of a 50 MHz clock. This timing requirement is inherited by all previous modules in the system. Our networking modules read in 1 byte every 4 cycles of the 50 MHz clock, and our audio module samples at a rate of 8 bytes per millisecond, meaning that we send an Ethernet packet every 0.1 seconds with 800 audio bytes. Similarly, the timing of the the video codec stack is governed by the \texttt{next\_packet} signal from the networking stack. The codec stack itself is pipelined appropriately to avoid timing issues. We used clock wizards as were necessary to generate clock signals for our input devices, codec, and networking stack.

\subsection{Project-Specific Evaluation}
\subsubsection{Compression}
We first evaluate compression on the raw compression ratio that we are able to achieve using the JPEG codec. At worst, the DCT coefficients for each pair of superblocks are 12 blocks of 64 11-bit values, which accounts for 1100 bytes. However, we find that the average payload size for video packets is 135 bytes, meaning we achieve an average compression ratio of 8 over the DCT coefficients, which is standard, but on the higher end for a JPEG-based implementation for small color representations and small images, which often struggle more with JPEG compression as the block size is nontrivial with respect to the size / features in the image itself.

We note that all our modules are testbenched against standardized numpy implementations and as such are numerically valid. When we reconstruct video, we can see the video as we expect, with motion captured qualitatively as we see in real life. However, we notice that because of bright sections of the image, because we already work with quantized values in the RGB space due to the 565 bit format, the DCT and quantization overly truncate signals and clamp them to 0 or 255. That being said, we note that this is standard for a MJPEG compression scheme with high compression ratio, which is necessary to stream video, with harsh quantization matrices. This has validated on reference implementations of MJPEG on small images similar to ours, where the aforementioned issues with small images also crop up.

\begin{figure}[thpb]
    \centering
    \includegraphics[width=\linewidth]{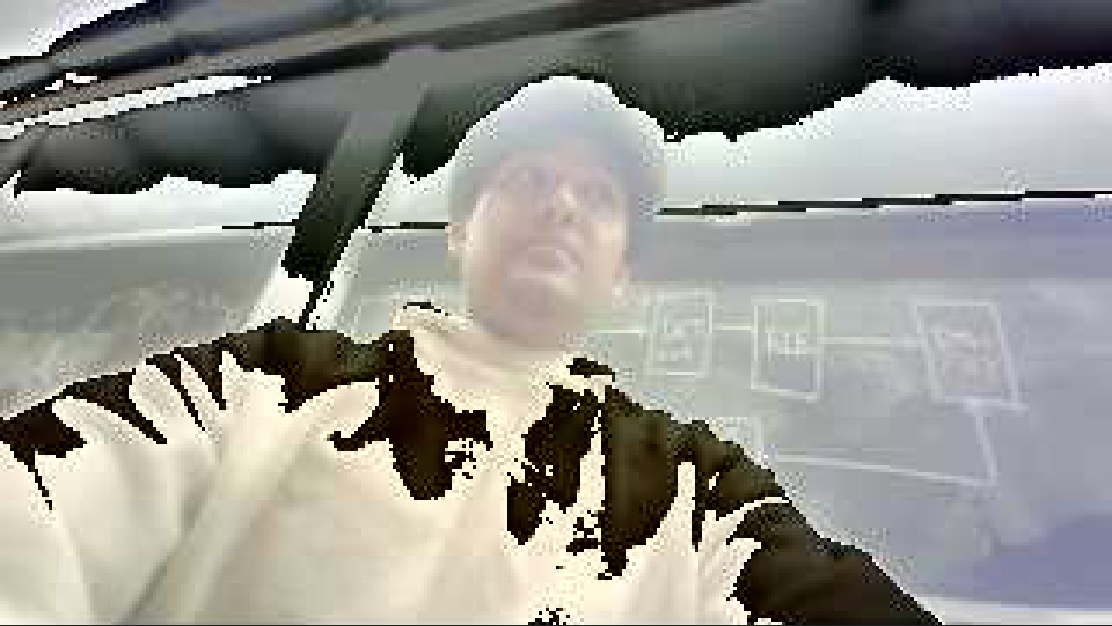}
    \caption{Image with Artifacts from JPEG compression}
    \label{fig:dct-artifacts}
\end{figure}

As such, we believe that our compression was implemented successfully and matches the MJPEG standard. In a future iteration of our system, we would love to experiment with higher resolution video and unquantized color signals, along with possibly modifying the codec / quantization schemes to better account for harsh lights and bright signals.

\subsubsection{Networking}
Firstly, we evaluate our networking performance. We observed 0 packet loss on streams up to 240,000 packets long, containing both video and audio data. Moreover, these packets are all in-order, as verified by the sequence metadata byte in their header. Because we sending our data over a single Ethernet cable to a constantly receiving client, such performance is optimal and expected. Our networking module is also fully testbenched, and we used Wireshark to verify data integrity over the ether.

On the audio streaming side, we can qualitatively evaluate our result after reconstruction. The audio stream is untouched by the Ethernet transmission, with it being of the quality we would expect directly from the board. Given that we perform no compression on this data, this is the expected outcome. The quality of the video feed is discussed above.

\subsubsection{Evaluation of Goals}
Our initial goal for this project was to create a fully functioning video codec and auxiliary networking stack to send compressed video over the network to a receiver, at 30 FPS. We were able to accomplish this goal fully, even achieving a much higher FPS of around 180. We were also able to implement Ethernet, IPv4, and UDP headers as desired, sending data using universal protocols. With that we have even reached our stretch goal of sreaming both audio and video over Ethernet using M-JPEG and UDP with throughput of at least 30 FPS. Having designed our modules with such ideals in mind, there are several ways we can improve the usage of our system.

The first is developing a better reconstruction script. Currently, our reconstruction runs on Python out of convenience, which means that it cannot sustain the throughput necessary for a live video stream. A next step would be to implement a more robust reconstruction algorithm, with better parallelism. This would also be much easier to accomplish on a non-MacOS machine, as this would enable us to interface directly with our machine's Ethernet port.

Another avenue of improvement is leveraging the multicast capability of UDP to send our video/audio stream to multiple recipient devices, thereby creating a more versatile conferencing setup. Because of our adherence to the UDP specification, such an improvement would not require much of a change in our design.

Finally, we could also build an FPGA-based receiver device to complete our on-chip conferencing setup. This receiver would have a simpler networking stack to just receive packets and store them, and an implementation of the inverse of our codec. It would write this video and audio reconstruction to a speaker and monitor.

\section{Implementation Insights}

While implementing the video codec, we had a tradeoff between using BRAM buffers to minimize register usage or serialization. After running into numerical bugs with the butterfly algorithm and 2D DCT FSM, we decided to go with serialization, as it was far easier to testbench with each output coming out one cycle at a time. On top of that, as we have sufficient time to serialize data without incurring too much dead time, as described in the Evaluation section, we chose to stick with a serialized ready-valid interface, as it was far easier to connect modules and test.

We also found an extreme quirk with using the OV5640 cameras on the Nexys4 DDR boards. We found that while clocking the camera xclk signal at 25 MHz worked on Urbana boards, the PMOD pins on the Nexys4 board do not support such a fast signal. Instead, we switch the xclk signal to a 12.5 MHz clock and the camera register assignment from 200 MHz to 100 MHz, and the cameras thus work with the Nexys4 DDR boards. However, we also encountered a new failure mode for cameras on this board. While the cameras would could overload some data pins on the Urbana board, when the Nexys4 board was not connected to external power, the cameras would inevitably burn out every port after a given time plugged in. We believe this may be a result of some kind of power cycling, but it is still unknown where it comes from.

While implementing the networking module, we encountered some interesting issues regarding packet malformation, extraneous response packets, and more. A key revelation was that the NIC (network interface card) of a MacBook Pro automatically discards packets with certain EtherTypes (notably, the IEEE experimental type we were using for testing), incorrect checksums, and invalidly sized packets. This nuance makes it somewhat difficult to debug issues with the Ethernet PHY, as we cannot pinpoint malformation in the packets. This became a major pain point during most of the development process. Moreover, MacOS automatically tries to resolve an IPv6 client through any Ethernet connection, through ARP and MDNS packets. In order to remove this, we had to properly configure our port to not send any such packets.

We also ran into some issues with how the Nexys DDR board sets up the LAN8720A PHY. In particular, the board uses an external pull up resistor on a configuration strap of the PHY, and then inverts the output of the LED that shares a pin with the strap. This unique configuration and other similar quirks make the behavior of the PHY somewhat distinct from that of the original spec. Luckily, we were able to use the Nexys 4 DDR schematic to clarify these items.

Finally, we also found that a lookup table implementation of the CRC-32 module was best for our system. At the expense of having a 8 by 255 bit lookup table that gets initialized into BRAM, we are able to support 1-cycle calculation of the CRC-32, under the configuration where we send it one byte of our packet at a time. This allows us to send full packets without having appended padding to give us time for checksum calculation.

\section{Conclusion}
Hence, in this project, we have developed, implemented, functionally verified, and synthesized a full pipeline for taking camera inputs, calculating an MJPEG codec, and transmitting over Ethernet, all on a Nexys4 DDR FPGA. We find that using proper pipelining and analysis of DSP / register load, we are easily able to meet timing at a 100 MHz clock and transmit 180p video at 180 FPS over a UDP connection, demonstrating the feasibility of FPGAs for video conferencing applications.



\end{document}